\documentstyle[12pt,aasms4]{article}
\def\gtsim{\lower.5ex\hbox{$\; \buildrel > \over \sim \;$}}
\def\ltsim{\lower.5ex\hbox{$\; \buildrel < \over \sim \;$}}

\slugcomment{}

\lefthead{Scharf \& Mushotzky} \righthead{Cluster Fe Abundances: A Measure
of Formation History ?}

\begin{document}

\title{ The Galaxy Cluster Luminosity-Temperature Relationship and
Iron Abundances:\\
 A Measure of Formation History ? }

\author{C.A. Scharf\altaffilmark{1} and R.F. Mushotzky}
\affil{Laboratory for High Energy Astrophysics, Code 662, NASA/Goddard
Space Flight Center, Greenbelt, MD 20771, USA}

\altaffiltext{1}{also, Dept. of Astronomy, University of Maryland, College
Park, MD 20742-2421, USA}

\begin{abstract} Both the X-ray luminosity-temperature (L-T) relationship
and the iron abundance distribution of galaxy clusters show intrinsic
dispersion. Using a large set of galaxy clusters with measured iron
abundances we find a correlation between abundance and the relative
deviation of a cluster from the mean L-T relationship. We argue that these
observations can be explained by taking into account the range of cluster
formation epochs expected within a hierarchical universe.  The known
relationship of cooling flow mass deposition rate to luminosity and
temperature is also consistent with this explanation.  From the observed
cluster population we estimate that the oldest clusters formed at $z\gtsim
2$. 

We propose that the iron abundance of a galaxy cluster can provide a
parameterization of its age and dynamical history. 

\end{abstract}


\section{Introduction}

The existence of an intrinsic spread or dispersion in the galaxy cluster
X-ray luminosity-temperature (L-T) relationship has been noted by several
authors (\cite{edg91}, \cite{fab94}, \cite{ms97} etc.).  Fabian et al.
(1994) demonstrated a correlation between the amplitude of the L-T
relationship and the cooling flow mass deposition rate
$\dot{M}$($M_{\odot}$yr$^{-1}$) ($L \propto T^{3.3} \dot{M}^{0.4}$). 

Recently Mushotzky \& Scharf (1997) have shown that the intrinsic
dispersion of the L-T relationship does not seem to evolve and remains
constant over a wide range in redshift ($z=0$ to $z=0.4$). Since the
advent of high precision cluster metallicity measurements (Yamashita et al
1992) it is also clear that there is a dispersion of a factor of 2 in
cluster Fe metallicities. This variation also does not evolve with
redshift (\cite{mus97}).  As Fabian et al (1994) have pointed out, it is
likely that temperatures, iron abundances and cooling flows are all linked
consequences of cluster histories. 

In this paper we show that the variance in the L-T relationship and the
cluster metallicity are correlated and can be explained in a simple model
of hierarchical clustering if the dispersion in the present L-T relation
reflects the {\em range} of cluster formation epochs. We propose that the
range in cluster formation epochs can simultaneously help explain the
correlation between the position of a cluster in the L-T relationship, its
metallicity, and the correlations with cooling flow rates.

\section{The L-T intrinsic dispersion}

Using a large sample (102 total, 39 in the luminosity range $45.2 \leq
\log_{10} L_{bol}\; ($erg s$^{-1}) \leq 45.7 $) of clusters we have
previously demonstrated (\cite{ms97}) that both the mean cluster
temperature and intrinsic temperature dispersion (at a fixed bolometric
luminosity) of the cluster population remains constant over the redshift
range $0.1 \ltsim z \ltsim 0.4$. Using likelihood analysis we have
estimated the intrinsic dispersion in the L-T relationship as
$\sigma_{T}\simeq 2$ keV at 7 keV, modeling the dispersion as a gaussian. 

We shall show that this intrinsic dispersion can be plausibly explained as
being largely due to a range of formation epochs in the cluster population
and thus the data can place some constraints on these epochs.  
Semi-analytic models of the cluster population typically assume that
the redshift of a cluster is approximately the redshift of its formation.
In an $\Omega=1$ universe dominated by cold dark matter this
is a justifiable simplification. It is not clear however that it should
apply to lower density cosmologies or to those with (for example) a
mixture of hot and cold dark matter. 
We use the
formalism of Kityama \& Suto (1997) who modify the Press-Schecter theory
to include the epoch of cluster formation as an explicit variable. Kityama
\& Suto suggest that the following is approximately true, assuming a self
similar model where the cluster core radius is proportional to the virial
radius;  
\begin{equation}
 T\propto M^{2/3} (1+z_{f})^{\xi} \left(\frac{1+z_{f}}{1+z}\right) ^{s}
\end{equation} 
\begin{equation} L_{bol} \propto M^{4/3}
(1+z_{f})^{\frac{7\xi}{2}} \left(\frac{1+z_{f}}{1+z}\right) ^{\frac{s}{2}} 
\end{equation} 
(c.f. \cite{evr91}) and ($z_{f}$) is the cluster
formation epoch and $\xi$ is an effective index that is unity for
$\Omega=1$ and no cosmological constant and varies weakly with $z_{f}$ for
low density cosmologies. A factor $((1+z_f)/(1+z))^s$ relates the 
observed cluster temperature at redshift $z$
to the virial temperature at $z_f$. Hydrodynamical simulations indicate 
that cluster
temperatures are consistent with $0\ltsim s \ltsim 1$ (e.g. \cite{nav95}).
Thus, at a given mass scale ($M$), clusters which formed at earlier epochs
are expected to be hotter and more luminous, with luminosity increasing
more rapidly than temperature.  Since the temperature evolution of
clusters is seen to be zero or small (\cite{ms97}) we expect $s\sim 0$. 

Preheating of the intracluster medium has been invoked (e.g. 
\cite{kai91}, \cite{evr91}) to reproduce the apparent {\em negative}
evolution seen in the luminosity function of the most
luminous clusters at high redshift (\cite{gio90}, \cite{hen92}), and to
better fit the locally observed L-T relation with semi-analytic models.
The assumptions made in such models (\cite{kai91}) imply that the heated
gas will contract until the gas temperature is roughly equal to the virial
temperature of the dark halo. The gas density profile is then set by the
cluster potential rather than the dark matter density contrast.  The
effect on Equations 1 \& 2 is that the overall $z_f$ dependence of the L-T
relation is weakened and therefore, for our purposes, this self-similar
case provides a lower limit in the determination of an effective $z_f$.

Using the measured temperature dispersion at a fixed luminosity
($\sigma_T\simeq 2$ keV at $45.2\leq \log_{10} L_{bol} \leq 45.7$), and
assuming a fixed mass scale, we can then proceed to calculate the
effective era of cluster formation. Combining Equations 1 \& 2 and
assuming that the observed $\sigma_T$ is due to the range of formation
epochs (from $z_{f}^{min}$ to $z_{f}^{max}$) we obtain;  $\log_{10}
((1+z_{f}^{max})/(1+z_{f}^{min}))=\frac{4}{3\xi}\log_{10} \sigma_T$, since
$s\sim 0$. 

For an $\Omega_0=1$ $(\lambda_o=0)$ universe ($\xi =1$) in which the most
recent cluster formation is at $z\sim 0$ then the earliest epoch of
cluster formation would be at $z\sim 1.5$ (with some at higher z, since we
model the L-T dispersion as a gaussian). Since lower density cosmologies
act to {\em increase} this upper bound we can make the general statement
that, if clusters are still forming at $z\sim 0$, then they must have
begun forming at $z\geq 1.5$. If the cluster population had essentially
finished forming by $z\sim 0.5$ then we would expect the earliest clusters
to have formed at $z\geq 2.5$. Given the lack of observed evolution in the
population of galaxy clusters to at least $z \sim 0.3$ (e.g. \cite{ebe97})
this latter result may be a better fit to observations.

\section{Abundance Data}

We have used a sample of 32 clusters ($0.01\ltsim z \ltsim 0.5$) with
precise Fe abundance measurements (averaged over the cluster),
temperatures and bolometric luminosities (\cite{mus97}, \cite{yam92} and
additional unpublished results) to determine if the Fe abundance is
correlated with the relative deviation of a cluster from the mean L-T 
relationship.

In Figure 1 we have plotted the observed Fe abundance relative to solar
(\cite{and89}) as a function of the amplitude of the L-T loci, hereafter
$A_{LT}$, assuming $L \propto T^3$, which has been determined to be a good
empirical fit (\cite{fab94}) as well as being close to theoretical
expectations (which range from $L\propto T^{2}$ to $L\propto T^{3.5}$).
The point size is proportional to the cluster temperature. 

The x-coordinate, $A_{LT}$, is therefore a direct measure of the `sense'
of the dispersion from a single power law fit to the L-T relation. The
larger $A_{LT}$ is, the more luminous a cluster is for a given
temperature. It is clear from Figure 1 that a correlation exists between
iron abundance and $A_{LT}$ such that clusters with higher abundances are
more luminous at fixed temperature.  The sole exception to this is the
Centaurus cluster which is the second closest luminous cluster and is
known to have a strong abundance gradient (\cite{fuk94}) and might
therefore be considered anomalous. A Spearman rank-order correlation test
on the data (with the Centaurus metallicity adjusted as indicated in
Figure 1) confirms that Fe and $A_{LT}$ are positively correlated with
98\% confidence and $r_s=0.43$. An unweighted least squares fit (plotted
in Figure 1) yields: $Fe\simeq 2.7\times 10^{-8} A_{LT}^{0.16}$. A maximum
likelihood analysis of the iron abundance residuals to this best fit
allows us to constrain the intrinsic dispersion (independent of $A_{LT}$)
in Fe to be $\leq 0.1$ (90\% confidence).  The cluster A2218, with the
lowest abundance in the sample, has been extensively studied (e.g.
\cite{squ96}) and is generally thought to be in an unrelaxed dynamical
state due to a recent large merger. There is no apparent correlation with
cluster temperature in the Fe-$A_{LT}$ plane. In Figure 2 the same data
are plotted, but with point size proportional to redshift, no redshift
correlation is seen in the Fe-$A_{LT}$ plane.

\section{Discussion}

From Equations 1 \& 2 the quantity $A_{LT}$ should increase with earlier
cluster formation epoch and thus we are led to interpret the data in
Figure 1 to mean that higher abundance clusters formed earlier. This has
profound implications for the nature of their formation. 

The metallicities observed in clusters are thought to have originated
primarily from SNe type II in an early epoch of star formation
(\cite{loe96}). In this interpretation the intracluster medium will have
undergone a significant 'pre-heating' at $z>1$ due to the energetics
required to match the observed abundances. Pre-heating is also favoured by
recent measurements of negligible, or slightly negative, cluster
luminosity evolution to $z\sim 0.3$ (e.g. \cite{ebe97}, \cite{jon97}) and
by measurements of negligible temperature evolution (e.g. \cite{ms97}). 
Recent semi-analytical models of galaxy formation (\cite{kau97}) indicate 
that 80\% of the intracluster metals were produced at $z>1$.

The sense of the relationship between $A_{LT}$ and abundance to $z_f$ can
be expressed in terms of a hierarchical `tree' history. In such a theory
there are two extremes. Either clusters have started as a single, dominant
potential well, into which much smaller clumps have fallen, or clusters
have formed by the merger of intermediate sized clumps over their history.
These two routes to a final cluster are, in fact, seen explicitly in halo
merger studies (e.g. \cite{lac93}, \cite{hus97}). We note that, for either
case, the rate of growth of the progenitor mass is generally slower for
low density cosmologies (\cite{hus97}). 

In the pre-heating scenario it is expected that low mass sub-clumps will,
because of their smaller potential wells, lose a substantial fraction of
their heated enriched gas during the epoch of metal formation while
massive systems will not. Thus a cluster formed later, whose evolution is
dominated by the merging of smaller sub-clumps, will tend to have a lower
metal abundance and a lower value of $A_{LT}$ than a massive system formed
earlier by the collapse of a single large overdensity.  This explanation
does however assume that clusters formed through these two routes have
roughly the same final virial mass.  In hierarchical models the most
massive systems are expected to form last so clearly this `direct'
interpretation of the L-T dispersion is a likely oversimplification of the
real play-off between system mass and $z_{f}$ in Equations 1 \& 2. 

Observational evidence (e.g. \cite{dav95}) indicates that (where
discernable) the subclumps involved in the formation of luminous clusters
at low redshift are themselves typically low luminosity systems, such as
groups. The abundances measured in groups are indeed often low
(\cite{fuk96}), consistent with our hypothesis.

Our interpretation of the correlation in Figure 1 is also consistent with
the results of Fabian et al (1994) which demonstrated that higher
$\dot{M}$ clusters have larger $A_{LT}$ and that higher $\dot{M}$ clusters
typically have higher abundances. Cooling flows are expected to be
disrupted by major mergers and hence objects which have not experienced
mergers are relatively undisturbed and, in the language used above, are
`truly' old. They would then also be expected to have higher $\dot {M}$'s. 
One of the `oldest' objects in our sample would then be 2A0335+096
(Figures 1 \& 2). There is evidence (\cite{irw95}) that this cluster is
indeed at a late stage of cooling, which places its age at $\gtsim 4$Gyr
(\cite{chr96}, and \cite{hu88}), or a $z_{f}$ of $\sim 2$, consistent with
the results in Section 2. 

It is important to note that the definitions of formation epochs (in terms
of mass halos) are actually somewhat arbitrary, for example Lacey \& Cole
(1993) propose a criteria that $z_f$ is the epoch at which a halo of mass
$M$ at $z$ had a mass greater than $M/2$ for the first time. Thus, the
above scenario is certainly overly simplistic and it is probably better to
consider cluster formation as a continuous process which, however, for
illustrative purposes can be subdivided into general classes such as the
two (realistic) extremes described above. 

Clearly, the range of abundances combined with the L-T dispersion results
shown above, should allow, through careful modeling, new constraints to be
placed on the overall pathways of hierarchical cluster formation.

\section{Conclusion}

We propose that the observed dispersion in the cluster L-T relationship
reflects the range in cluster formation epochs within a hierarchical
universe. The relationship of cooling flow mass deposition with luminosity
and temperature is then qualitatively explained if we assume that cooling
flow clusters are both initially denser and older than non-cooling flow
clusters, which we associate with recent major mergers. The observed
dispersion in cluster iron abundances and its correlation with position in
the L-T diagram ($A_{LT}$) is accounted for (except for a possible
residual intrinsic dispersion of $\Delta Fe\leq 0.1$) if the smaller mass
units that are involved in the  mergers can lose metals via an early
cluster wind phase and then merge to form the lower $A_{LT}$ systems.  In
this scenario we can constrain the earliest cluster formation to be at
$z_f \gtsim 2$.  While there are clearly other candidate mechanisms for
producing the observed variances in abundances and temperatures (e.g.
inhomogeneity of the intracluster medium (\cite{fab94}), these must
ultimately also be functions of the cluster age and history. 

Larger data sets and more detailed cosmological simulations will greatly
improve our understanding of the results presented here and will test our
hypothesis that the measurement of a clusters iron abundance informs us of
its age and formation history. For example, we expect that higher
abundance clusters will typically have higher baryonic fractions. We also
expect that the most massive clusters observed at $z\simeq 1-2$ will
typically have higher abundances than their local counterparts. In future
work we will investigate these predictions.

\clearpage 

\figcaption[]{Measured iron abundance (open circles) is plotted against
$L/(T^3)$ (as the logarithm), the amplitude of the L-T power law required
for a given cluster: $A_{LT}$. Circle size is linearly proportional to
cluster temperature. The solid line is an unweighted least squares fit to
the data and indicates the sense of the proposed abundance relationship. 
Virgo, Centaurus, A2218, A1060, A1204 and 2A0335+096 are labelled. The
Centaurus cluster Fe abundance (from Ginga) is considered anomalous due to
strong abundance gradients, the vertical arrow indicates the correction
obtained by going from emission measure weighted abundance to average
abundance.}

\figcaption[]{As for Figure 1, but point size is now linearly proportional
to cluster redshift}

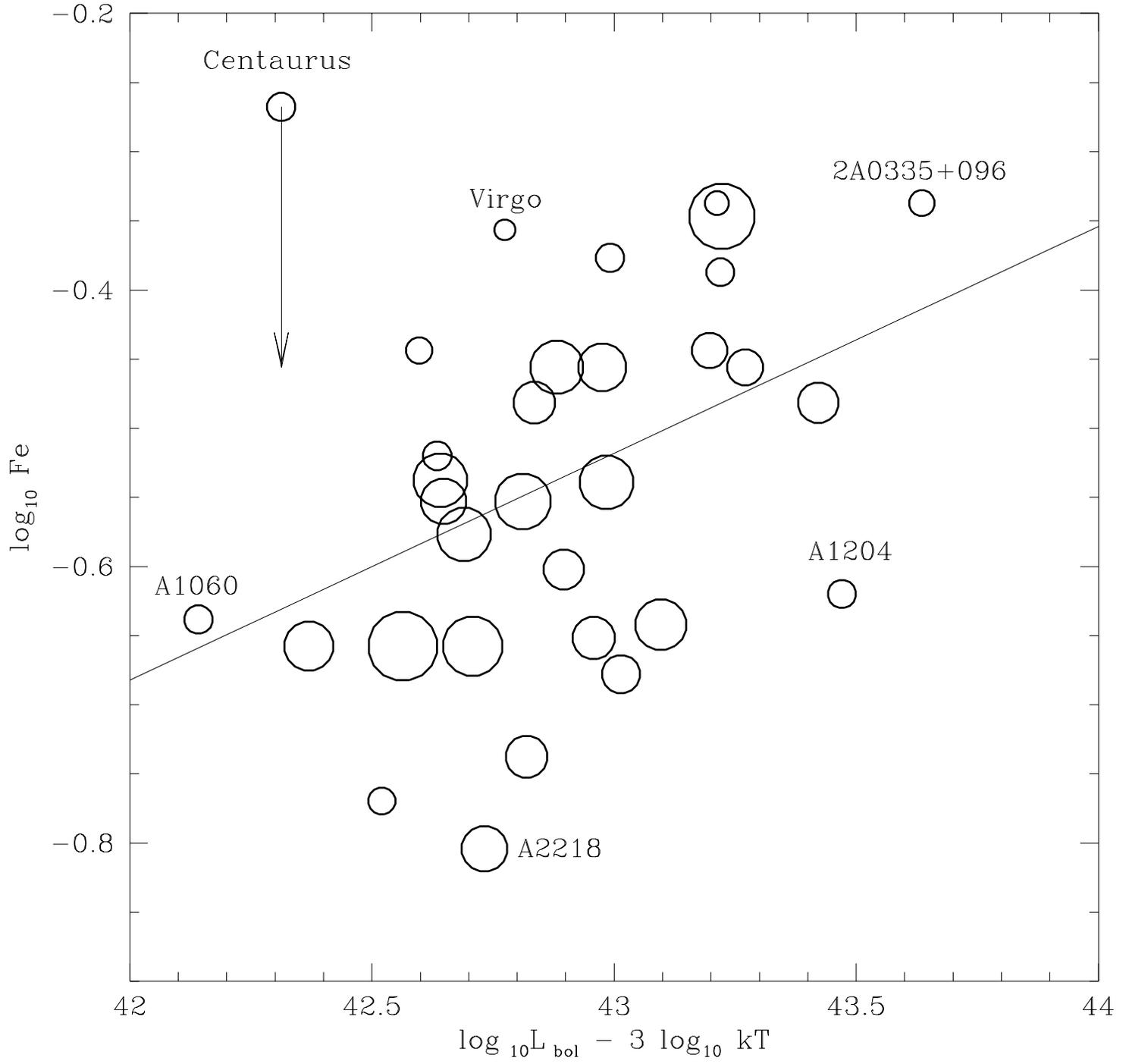
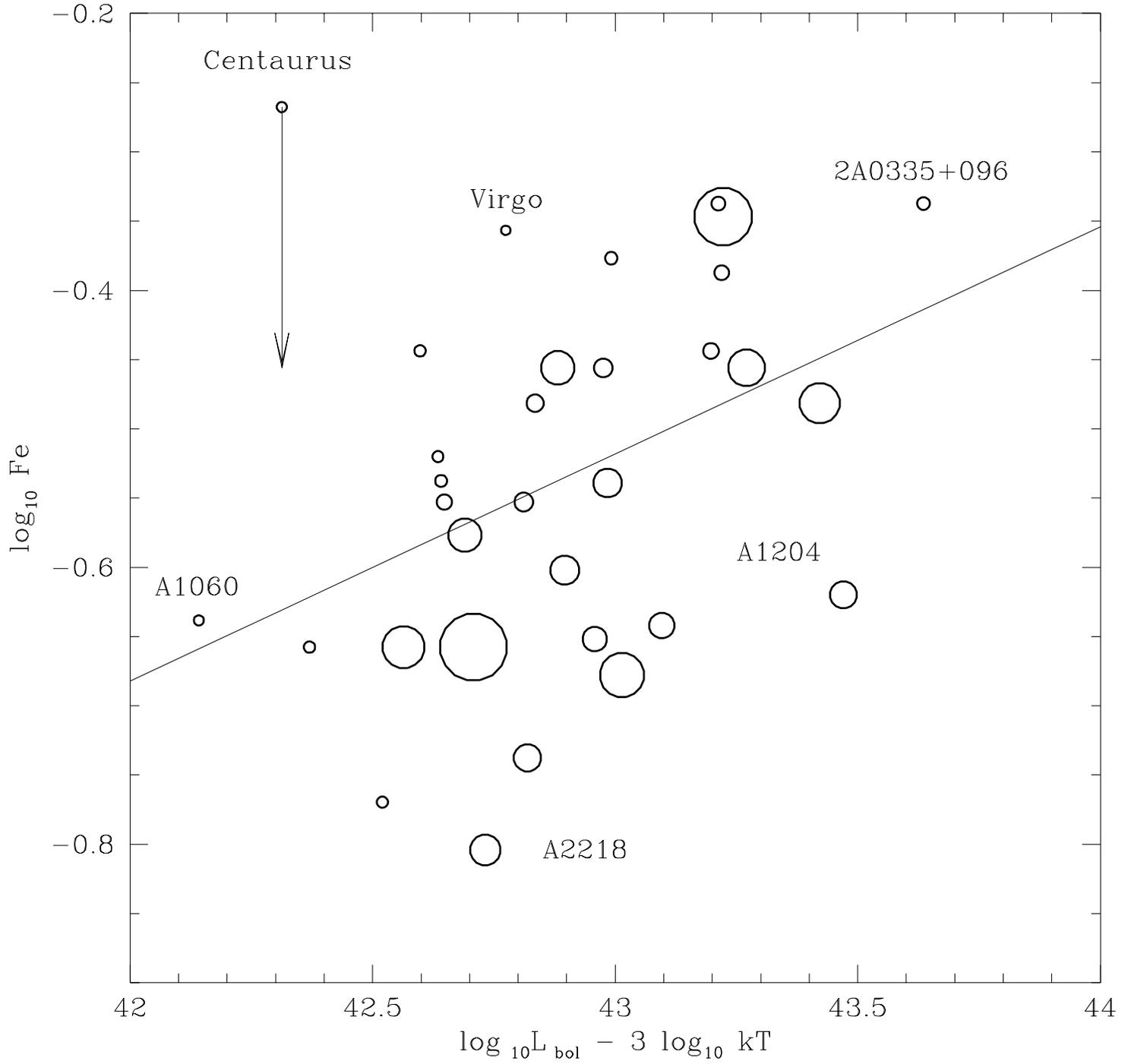
\end{document}